# A new look at the problem of gauge invariance in quantum field theory

by

Dan Solomon

Rauland-Borg Corporation
3450 W. Oakton Street
Skokie, IL 60076
USA

Phone: 1-847-324-8337
Email: dan.Solomon@rauland.com

## Abstract

Quantum field theory is assumed to be gauge invariant. However it is well known that when certain quantities are calculated using perturbation theory the results are not gauge invariant. The non-gauge invariant terms have to be removed in order to obtain a physically correct result. In this paper we will examine this problem and determine why a theory that is supposed to be gauge invariant produces non-gauge invariant results.

## 1. Introduction

It is well know that quantum field theory contains anomalies. An anomaly occurs when the result of a calculation does not agree with some underlying symmetry of the theory. Such is the case with gauge invariance.

Quantum field theory is assumed to be gauge invariant [1][2]. A change in the gauge is a change in the electromagnetic potential that does not produce a change in the electromagnetic field. The electromagnetic field is given by,

$$\vec{E} = -\left(\frac{\partial \vec{A}}{\partial t} + \vec{\nabla} A_0\right); \quad \vec{B} = \vec{\nabla} \times \vec{A} \tag{1.1}$$

where $\vec{E}$ is the electric field, $\vec{B}$ is the magnetic field, and $\left(A_0, \vec{A}\right)$ is the electromagnetic potential. A change in the electromagnetic potential that does not produce a change the electromagnetic field is given by,

$$\vec{A} \to \vec{A}' = \vec{A} - \vec{\nabla}\chi, \quad A_0 \to A_0' = A_0 + \frac{\partial \chi}{\partial t} \tag{1.2}$$

where $\chi(\vec{x}, t)$ is an arbitrary real valued function. Using relativistic notation this can also be written as,

$$A_\nu \to A_\nu' = A_\nu + \partial_\nu \chi \tag{1.3}$$

In order for quantum field theory to be gauge invariant a change in the gauge cannot produce a change in any physical observable such as the current and charge expectation values. However it is well known that when certain quantities are calculated using standard perturbation theory the results are not gauge invariant. The non-gauge invariant terms that appear in the results have to be removed to make the answer physically correct.

For example, the first order change in the vacuum current, due to an applied electromagnetic field, can be shown to be given by,

$$J^{\mu}_{vac}(x) = \int \pi^{\mu\nu}(x - x') A_{\nu}(x') d^4x' \tag{1.4}$$

where $\pi^{\mu\nu}$ is called the polarization tensor and where, in the above expression, summation over repeated indices is assumed. The above equation is normally written in terms of the Fourier transformed quantities as,

$$J^{\mu}_{vac}(k) = \pi^{\mu\nu}(k) A_{\nu}(k) \tag{1.5}$$

where k is the 4-momentum of the electromagnetic field. In this case a gauge transformation takes the following form,

$$A_{\nu}(k) \rightarrow A'_{\nu}(k) = A_{\nu}(k) + ik_{\nu}\chi(k) \tag{1.6}$$

The change in the vacuum current, $\delta_g J^{\mu}_{vac}(k)$, due to a gauge transformation can be obtained by using (1.6) in (1.5) to yield,

$$\delta_g J^{\mu}_{vac}(k) = ik_{\nu}\pi^{\mu\nu}(k)\chi(k) \tag{1.7}$$

Now if quantum theory is gauge invariant then an observable quantity, such as the vacuum current, must not be affected by a gauge transformation. Therefore $\delta_g J^{\mu}_{vac}(k)$ must be zero. For this to be true we must have that,

$$k_{\nu}\pi^{\mu\nu}(k) = 0 \tag{1.8}$$

However, when the polarization tensor is calculated it is found that the above relationship does not hold.

Consider, for example, a calculation of the vacuum polarization tensor by Heitler (see page 322 of [3]). Heitler's solution for the Fourier transform of the vacuum polarization tensor is,

$$\pi^{\mu\nu}(k) = \pi_G^{\mu\nu}(k) + \pi_{NG}^{\mu\nu}(k) \tag{1.9}$$

The first term on the right hand side is given by,

$$\pi_G^{\mu\nu}(k) = \left(\frac{2q^2}{3\pi}\right)\left(k^\mu k^\nu - g^{\mu\nu}k^2\right)\int_{2m}^{\infty} dz \frac{\left(z^2 + 2m^2\right)\sqrt{\left(z^2 - 4m^2\right)}}{z^2\left(z^2 - k^2\right)} \tag{1.10}$$

where m is the mass of the electron, q is the electric charge, and $\hbar = c = 1$. This term is gauge invariant because $k_\nu \pi_G^{\mu\nu} = 0$. The second term on the right of (1.9) is

$$\pi_{NG}^{\mu\nu}(k) = \left(\frac{2q^2}{3\pi}\right) g_\nu^\mu \left(1 - g^{\mu 0}\right)\int_{2m}^{\infty} dz \frac{\left(z^2 + 2m^2\right)\sqrt{\left(z^2 - 4m^2\right)}}{z^2} \tag{1.11}$$

where there is no summation over the two $\mu$ superscripts that appear on the right. Note that $\pi_{NG}^{\mu\nu}$ is not gauge invariant because $k_\nu \pi_{NG}^{\mu\nu} \neq 0$. Therefore to get a physically valid result it is necessary to "correct" equation (1.9) by dropping $\pi_{NG}^{\mu\nu}$ from the solution.

A similar situation exists when other sources in the literature are examined. For example consider the discussion in Section 14.2 of Greiner et al [2]. Greiner writes the solution for the vacuum polarization tensor (see equation 14.43 of [2]) as,

$$\pi^{\mu\nu}(k) = \left(g^{\mu\nu}k^2 - k^\mu k^\nu\right)\pi\left(k^2\right) + g^{\mu\nu}\pi_{sp}\left(k^2\right) \tag{1.12}$$

where the quantities $\pi\left(k^2\right)$ and $\pi_{sp}\left(k^2\right)$ are given in [2]. Referring to (1.8) it can be easily shown that the first term on the right is gauge invariant. However the second term is not gauge invariant unless $\pi_{sp}\left(k^2\right)$ equals zero. Greiner shows that this is not the case. Therefore this term must be dropped form the result in order to obtain a physically valid solution.

For another example of this refer to section 6-4 of Nishijima [4]. In this reference is it shown that the vacuum polarization tensor includes a non-gauge invariant term which must be removed. For other examples refer to equation 7.79 of Peskin and Schroeder [5] and Section 5.2 of Greiner and Reinhardt [6]. In all cases a direct calculation of the vacuum polarization tensor using perturbation theory produces a result which includes non-gauge invariant terms. In all cases the non-gauge invariant terms must be removed to obtain the "correct" gauge invariant result.

There are two general approaches to removing these non-gauge invariant terms. The first approach is simply to recognize that the term cannot be physically valid and drop it from the solution. This is the approach taken by Heitler [3], Nishijima [4], and Greiner et al [2]. The other approach is to come up with mathematical techniques which automatically eliminate the offending terms. This is called "regularization". One type of regularization is called Pauli-Villars regularization [7] . In this case additional functions are introduced that have the correct behavior so that the non-gauge invariant terms are cancelled. An example of the use Pauli-Villars regularization is given by Greiner and Reinhardt [6]. Another type of regularization is called dimensional regularization. An example of this is given by Peskin and Schroeder [5].

The problem with regularization is that there is no physical explanation for why it is required. It is an ad hoc mathematical device required to remove the unwanted terms. Also regularization does not guarantee a unique answer. For example consider the result from [2] given by equation (1.12). As discussed above the second term on the right must be removed. In reference [2] the authors simply removed the term $g^{\mu\nu}\pi_{sp}\left(k^2\right)$ by hand without resorting to formal regularization schemes. Suppose someone invented a mathematical procedure that subtracted the term $g^{\mu\nu}\pi_{sp}\left(k^2\right)+\left(g^{\mu\nu}k^2-k^{\mu}k^{\nu}\right)f\left(k^2\right)$ from (1.12) where $f\left(k^2\right)$ is some arbitrary function. The final "corrected" result would be $\left(g^{\mu\nu}k^2-k^{\mu}k^{\nu}\right)\left(\pi\left(k^2\right)-f\left(k^2\right)\right)$. Since this is gauge invariant it would certainly be a physically acceptable solution, however it is not unique because $f\left(k^2\right)$ is arbitrary. That is, there is no theoretical way to distinguish between a mathematical regularization procedure for which $f\left(k^2\right)=0$ and one for which $f\left(k^2\right)$ is non-zero. Both yield physically valid results.

The obvious question that should be asked is why do non-gauge invariant terms appear in a theory that is supposed to be gauge invariant? This question is briefly addressed by Greiner et al (page 398 of [2]) who writes "… this latter [non-gauge invariant] term violates the gauge invariance of the theory. This is a very sever contradiction to the experimentally confirmed gauge independence of QED. [This problem indicates] that perturbative QED is not a complete theory. As one counter example or inconsistency suffices to prove a theory wrong, we should, in principle, spend the rest of this book searching for an improved theory. However, there is little active work on this today because: (1) there is a common belief that some artifact of the exact mathematics is the source of the problem; (2) this problem may disappear when the properly generalized theory, including in its framework all charged Dirac particles, is achieved."

It is my impression that the above paragraph pretty much expresses the current state of thinking on this problem, i.e. , the problem is probably due to an artifact of the mathematics and will, hopefully, go away when a complete theory is revealed. However recent work comparing Dirac's hole theory (HT) to quantum field theory (QFT) suggests that we consider the problem from a different perspective.

Consider a "simple" quantum theory consisting of non-interacting electrons in a background classical electromagnetic field. For such a situation Dirac's hole theory and quantum field theory are generally assumed to be equivalent. Hole theory was introduced by Dirac to resolve the problem caused by the fact that solutions of the Dirac equation include both positive and negative energy solutions. Dirac proposed that all the negative energy states are occupied by a single electron and then evoked the Pauli exclusion principle to prevent the decay of a positive energy electron into negative energy states. The electrons in the negative energy states, the so called *Dirac sea*, are assumed to be unobservable. What we observe are variations from the unperturbed vacuum state.

Recently several papers have appeared in the literature pointing out that there are differences between Dirac's hole theory (HT) and quantum field theory (QFT) [8][9][10][11][12][13]. The problem was originally examined by Coutinho et al[8][9]. They calculated the second order change in the energy of the vacuum state due to a time independent perturbation. They found that HT and QFT produce different results. They concluded that the difference between HT and QFT was related to the validity of Feynman's belief that the Pauli Exclusion Principle can be disregarded for intermediate states in perturbation theory. This belief was based on Feynman's observation that terms that violate the Pauli principle formally cancel out in perturbation theory. However Coutino et al show that this is not necessarily the case for HT when applied to an actual problem. One interesting result of this research was pointed out by the author(Solomon [13] ) where it was shown that HT is anomaly free. Therefore it may be possible to understand why QFT contains anomalies by examining the differences between HT and QFT.

One such difference was described in [13] and [14]. In these papers it was shown that the HT vacuum state is not a state of minimum energy. That is, there exist states with less energy than the energy of the vacuum in HT. This is a somewhat surprising result because it is generally assumed that the HT vacuum state is also a minimum energy state. This is in sharp contrast to QFT where the vacuum can easily be shown to be the minimum energy state. It will be shown that this difference is the reason for the gauge invariance anomaly in QFT.

Therefore it is the purpose of this paper to more fully address the question on why non-gauge invariant terms appear in calculations of the vacuum current in QFT. It will be shown that this is not an artifact of the mathematics but is a result of the underlying structure of QFT and, in particular, the properties of the vacuum state.

The discussion will proceed as follows. In Section 2 some basic elements of formal field theory will be introduced. It will then be shown in Section 3 that these elements are not mathematically consistent and that this inconsistency is related to the requirement of gauge invariance. In Section 4 the change in the vacuum current as a result of a gauge transformation will be calculated. This calculation is performed in 1-1D space-time. The advantage of this is that all integrals are finite and well defined. Therefore there is no need to regularize divergent integrals. It is shown that the results are not gauge invariant.

**2. Quantum Field theory.**

In this section the basic elements of quantum field theory in the Schrödinger representation will be introduced. Natural unit will be used so that $\hbar = c = 1$. We shall consider a "simple" field theory consisting of non-interacting fermions acted on by a classical electromagnetic field. In the Schrödinger representation of QFT the time evolution of the state vector $\left|\Omega(t)\right\rangle$ and its dual $\left\langle\Omega(t)\right|$ are given by,

$$\frac{\partial\left|\Omega(t)\right\rangle}{\partial t} = -i\hat{H}(t)\left|\Omega(t)\right\rangle, \quad \frac{\partial\left\langle\Omega(t)\right|}{\partial t} = i\left\langle\Omega(t)\right|\hat{H}(t) \tag{2.1}$$

where $\hat{H}(t)$ is the Hamiltonian operator which is given by,

$$\hat{H}(t) = \hat{H}_0 - \int \hat{\vec{J}}(\vec{x}) \cdot \vec{A}(\vec{x},t) d\vec{x} + \int \hat{\rho}(\vec{x}) A_0(\vec{x},t) d\vec{x} \tag{2.2}$$

In the above expression the quantities $\left(A_0(\vec{x},t), \vec{A}(\vec{x},t)\right)$ are the electromagnetic potential which are assumed to be unquantized, real valued functions. The quantity $\hat{H}_0$ is the free field Hamiltonian, that is, the Hamiltonian when the electromagnetic potential is zero, and $\hat{\vec{J}}(\vec{x})$ and $\hat{\rho}(\vec{x})$ are the current and charge operators, respectively. Note that $\hat{H}_0$, along with the current and charge operators, are time independent which is consistent with the Schrödinger picture approach. Throughout this discussion it is assumed that $\left|\Omega(t)\right\rangle$ is normalized, i.e., $\left\langle\Omega(t)\middle|\Omega(t)\right\rangle = 1$. Note that Eq. (2.1) ensures that the normalization of $\left|\Omega(t)\right\rangle$ is constant in time.

A second element of QFT is the principal of gauge invariance which was discussed in the previous section. When a change in the gauge is introduced into (2.1) this will produce a change in the state vector $|\Omega\rangle$. However, for QFT to be gauge invariant a change in the gauge must produce no change in the physical observables. These include the current and charge expectation values which are defined by,

$$\vec{J}_e(\vec{x},t) = \langle\Omega(t)|\hat{\vec{J}}(\vec{x})|\Omega(t)\rangle \text{ and } \rho_e(\vec{x},t) = \langle\Omega(t)|\hat{\rho}(\vec{x})|\Omega(t)\rangle \quad (2.3)$$

A third element that we expect QFT to obey is that of local conservation of electric charge, that is, the continuity equation holds,

$$\frac{\partial\rho_e(\vec{x},t)}{\partial t} + \vec{\nabla}\cdot\vec{J}_e(\vec{x},t) = 0 \quad (2.4)$$

The fourth element of Dirac theory to be considered in this discussion is that there exists a minimum value to the free field energy. The free field energy, $\xi_0(|\Omega\rangle)$, of a normalized state vector $|\Omega\rangle$, is the energy of the quantum state in the absence of interactions, i.e., the electromagnetic potential is zero. $\xi_0(|\Omega\rangle)$ is defined by,

$$\xi_0(|\Omega\rangle) = \langle\Omega|\hat{H}_0|\Omega\rangle \quad (2.5)$$

Let $|n\rangle$ be the eigenstates of $\hat{H}_0$ with eigenvalues $\varepsilon_n$. The $|n\rangle$ form a complete orthonormal set of basis states and satisfy the equations,

$$\hat{H}_o|n\rangle = \varepsilon_n|n\rangle; \quad \langle n|\hat{H}_o = \langle n|\varepsilon_n \quad (2.6)$$

and,

$$\langle n|m\rangle = \delta_{nm} \quad (2.7)$$

and,

$$\sum_n |n\rangle\langle n| = \hat{1} \qquad (2.8)$$

Any arbitrary state $|\Omega\rangle$ can be expanded in terms of the eigenstates $|n\rangle$ so that we can write,

$$|\Omega\rangle = \sum_n c_n |n\rangle \qquad (2.9)$$

where $c_n$ are the expansion coefficients.

The vacuum state $|0\rangle$ is generally assumed to be the eigenvector of $\hat{H}_0$ with the smallest eigenvalue $\varepsilon_o = 0$. For all other eigenvalues,

$$\varepsilon_n > \varepsilon_o = 0 \text{ for } |n\rangle \neq |0\rangle \qquad (2.10)$$

Using this fact along with (2.5)-(2.9) we can easily show that,

$$\xi_0\left(|\Omega\rangle\right) > \xi_0\left(|0\rangle\right) = 0 \text{ for all } |\Omega\rangle \neq |0\rangle \qquad (2.11)$$

Therefore the vacuum state is the quantum state with the minimum value of the free field energy.

## **3. A mathematical inconsistency.**

The four elements of QFT field theory that were introduced in the last section were (1)the Schrodinger equation, (2)the principle of gauge invariance, (3)the continuity equation, and (4)the idea that the vacuum state $|0\rangle$ is the state with minimum free field energy. The question that we will address in this section is whether or not these four elements of Dirac field theory are mathematically consistent. From the equations of Section 2 we will derive a number of additional relationships. First, consider the time derivative of the current expectation value. From (2.3) and (2.1) we obtain,

$$\frac{\partial \vec{J}_e(\vec{x},t)}{\partial t} = i\left\langle \Omega(t) \left| \left[ \hat{H}(t), \hat{\vec{J}}(\vec{x}) \right] \right| \Omega(t) \right\rangle \tag{3.1}$$

Use (2.2) in the above to yield,

$$\frac{\partial \vec{J}_e(\vec{x},t)}{\partial t} = i\left\langle \Omega(t) \left| \begin{pmatrix} \left[ \hat{H}_0, \hat{\vec{J}}(\vec{x}) \right] - \int \left[ \hat{\vec{J}}(\vec{y}) \cdot \vec{A}(\vec{y},t), \hat{\vec{J}}(\vec{x}) \right] d\vec{y} \\ + \int \left[ \hat{\rho}(\vec{y}), \hat{\vec{J}}(\vec{x}) \right] A_0(\vec{y},t)\, d\vec{y} \end{pmatrix} \right| \Omega(t) \right\rangle \tag{3.2}$$

Next perform the gauge transformation (1.2) to obtain,

$$\frac{\partial \vec{J}_e(\vec{x},t)}{\partial t} = i\left\langle \Omega(t) \left| \begin{pmatrix} \left[ \hat{H}_0, \hat{\vec{J}}(\vec{x}) \right] - \int \left[ \hat{\vec{J}}(\vec{y}) \cdot \left( \vec{A}(\vec{y},t) - \vec{\nabla}\chi(\vec{y},t) \right), \hat{\vec{J}}(\vec{x}) \right] d\vec{y} \\ + \int \left[ \hat{\rho}(\vec{y}), \hat{\vec{J}}(\vec{x}) \right] \left( A_0(\vec{y},t) + \frac{\partial \chi(\vec{y},t)}{\partial t} \right) d\vec{y} \end{pmatrix} \right| \Omega(t) \right\rangle \tag{3.3}$$

The quantity $\partial \vec{J}_e / \partial t$ is a physical observable and therefore, if the theory is gauge invariant, must not depend on the quantity $\chi$ or $\partial \chi / \partial t$. Now, at a particular instant of time $\partial \chi / \partial t$ can be varied in an arbitrary manner without changing the values of any of the other quantities on the right hand side of the equals sign in the above equation. Therefore for $\partial \vec{J}_e / \partial t$ to be independent of $\partial \chi / \partial t$ we must have that,

$$ST(\vec{y},\vec{x}) \equiv \left[ \hat{\rho}(\vec{y}), \hat{\vec{J}}(\vec{x}) \right] = 0 \tag{3.4}$$

We will call $ST(\vec{y},\vec{x})$ the Schwinger term. Next use (2.3) in (2.4) to obtain,

$$\vec{\nabla} \cdot \vec{J}_e(\vec{x},t) = -\frac{\partial \left\langle \Omega(t) \left| \hat{\rho}(\vec{x}) \right| \Omega(t) \right\rangle}{\partial t} \tag{3.5}$$

Next use (2.1) in the above to yield,

$$\vec{\nabla} \cdot \vec{J}_e(\vec{x},t) = -i\left\langle \Omega(t) \left| \left[ \hat{H}, \hat{\rho}(\vec{x}) \right] \right| \Omega(t) \right\rangle \tag{3.6}$$

Use (2.2) in the above to yield,

$$\vec{\nabla}\cdot\vec{J}_e(\vec{x},t) = -i\left\langle\Omega(t)\right|\begin{pmatrix}\left[\hat{H}_0,\hat{\rho}(\vec{x})\right]-\int\left[\hat{\vec{J}}(\vec{y}),\hat{\rho}(\vec{x})\right]\cdot\vec{A}(\vec{y},t)d\vec{y} \\ +\int\left[\hat{\rho}(\vec{y}),\hat{\rho}(\vec{x})\right]A_0(\vec{y},t)d\vec{y}\end{pmatrix}\left|\Omega(t)\right\rangle \tag{3.7}$$

Use (3.4) in the above to obtain,

$$\vec{\nabla}\cdot\vec{J}_e(\vec{x},t) = -i\left\langle\Omega(t)\right|\left(\left[\hat{H}_0,\hat{\rho}(\vec{x})\right]+\int\left[\hat{\rho}(\vec{y}),\hat{\rho}(\vec{x})\right]A_0(\vec{y},t)d\vec{y}\right)\left|\Omega(t)\right\rangle \tag{3.8}$$

We can then apply a gauge transformation to obtain,

$$\vec{\nabla}\cdot\vec{J}_e(\vec{x},t) = -i\left\langle\Omega(t)\right|\left(\left[\hat{H}_0,\hat{\rho}(\vec{x})\right]+\int\left[\hat{\rho}(\vec{y}),\hat{\rho}(\vec{x})\right]\begin{pmatrix}A_0(\vec{y},t) \\ +\dfrac{\partial\chi(\vec{y},t)}{\partial t}\end{pmatrix}d\vec{y}\right)\left|\Omega(t)\right\rangle \tag{3.9}$$

The quantity $\vec{\nabla}\cdot\vec{J}_e(\vec{x},t)$ is an observable and must be independent of $\partial\chi/\partial t$. Therefore,

$$\left[\hat{\rho}(\vec{y}),\hat{\rho}(\vec{x})\right]=0 \tag{3.10}$$

so that,

$$\vec{\nabla}\cdot\vec{J}_e(\vec{x},t) = \left\langle\Omega(t)\right|\vec{\nabla}\cdot\hat{\vec{J}}(\vec{x})\left|\Omega(t)\right\rangle = -i\left\langle\Omega(t)\right|\left[\hat{H}_0,\hat{\rho}(\vec{x})\right]\left|\Omega(t)\right\rangle \tag{3.11}$$

In order for the above equation to be true for arbitrary values of the state vector $\left|\Omega(t)\right\rangle$ we must have,

$$i\left[\hat{H}_0,\hat{\rho}(\vec{x})\right] = -\vec{\nabla}\cdot\hat{\vec{J}}(\vec{x}) \tag{3.12}$$

Now consider relationships (3.4) and (3.12). They following directly from the first three of the four elements of Dirac field theory that we have introduced in Section 1. However Schwinger [15] has shown that they are not compatible with the fourth element, that is, the assumption that the vacuum state is the state with the lowest free field energy. This will be demonstrated below.

First take the divergence of the Schwinger term $\left[\hat{\rho}(\vec{y}),\hat{J}(\vec{x})\right]$ and use (3.12) to obtain,

$$\vec{\nabla}_{\vec{x}}\cdot\left[\hat{\rho}(\vec{y}),\hat{\vec{J}}(\vec{x})\right]=\left[\hat{\rho}(\vec{y}),\vec{\nabla}\cdot\hat{\vec{J}}(\vec{x})\right]=-i\left[\hat{\rho}(\vec{y}),\left[\hat{H}_0,\hat{\rho}(\vec{x})\right]\right] \quad (3.13)$$

Next expand the commutator to yield,

$$i\vec{\nabla}_{\vec{x}}\cdot\left[\hat{\rho}(\vec{y}),\hat{\vec{J}}(\vec{x})\right]=-\hat{H}_0\hat{\rho}(\vec{x})\hat{\rho}(\vec{y})+\hat{\rho}(\vec{x})\hat{H}_0\hat{\rho}(\vec{y})+\hat{\rho}(\vec{y})\hat{H}_0\hat{\rho}(\vec{x})-\hat{\rho}(\vec{y})\hat{\rho}(\vec{x})\hat{H}_0 \quad (3.14)$$

Sandwich the above expression between the state vector $\langle 0|$ and its dual $|0\rangle$ and use $\hat{H}_0|0\rangle=0$ and $\langle 0|\hat{H}_0=0$ to obtain,

$$i\vec{\nabla}_{\vec{x}}\cdot\langle 0|\left[\hat{\rho}(\vec{y}),\hat{\vec{J}}(\vec{x})\right]|0\rangle=\langle 0|\hat{\rho}(\vec{x})\hat{H}_0\hat{\rho}(\vec{y})|0\rangle+\langle 0|\hat{\rho}(\vec{y})\hat{H}_0\hat{\rho}(\vec{x})|0\rangle \quad (3.15)$$

Next set $\vec{y}=\vec{x}$ to obtain,

$$i\vec{\nabla}_{\vec{x}}\cdot\langle 0|\left[\hat{\rho}(\vec{y}),\hat{\vec{J}}(\vec{x})\right]|0\rangle\underset{\vec{y}=\vec{x}}{=}2\langle 0|\hat{\rho}(\vec{x})\hat{H}_0\hat{\rho}(\vec{x})|0\rangle \quad (3.16)$$

Use (2.8) in the above to obtain,

$$i\vec{\nabla}_{\vec{x}}\cdot\langle 0|\left[\hat{\rho}(\vec{y}),\hat{\vec{J}}(\vec{x})\right]|0\rangle\underset{\vec{y}=\vec{x}}{=}2\sum_{n,m}\langle 0|\hat{\rho}(\vec{x})|n\rangle\langle n|\hat{H}_0|m\rangle\langle m|\hat{\rho}(\vec{x})|0\rangle \quad (3.17)$$

Next use (2.6) and (2.7) to obtain,

$$i\vec{\nabla}_{\vec{x}}\cdot\langle 0|\left[\hat{\rho}(\vec{y}),\hat{\vec{J}}(\vec{x})\right]|0\rangle\underset{\vec{y}=\vec{x}}{=}2\sum_n\varepsilon_n\langle 0|\hat{\rho}(\vec{x})|n\rangle\langle n|\hat{\rho}(\vec{x})|0\rangle=2\sum_n\varepsilon_n\left|\langle 0|\hat{\rho}(\vec{x})|n\rangle\right|^2 \quad (3.18)$$

Now, in general, the quantity $\langle 0|\hat{\rho}(\vec{x})|n\rangle$ is not zero [15] and since $\varepsilon_n\geq 0$ (see equation (2.10)) the above expression is non-zero and positive. Therefore the Schwinger term cannot be zero. This is, of course, in direct contradiction to (3.4).

Therefore there is a mathematical inconsistency in the theory. If the condition given by (2.10) is true then the Schwinger term is not zero. However in order for the theory to be gauge invariant the Schwinger term must be zero. This can be seen from examining (3.3). If the Schwinger term is not zero then the observable $\partial\vec{J}_e/\partial t$ is dependent on the gauge transformation. The result of all this is that if (2.10) is true then the theory is not, in fact, gauge invariant. This will be shown in the next section where a calculation of the vacuum current will be shown to be non-gauge invariant.

**4. A calculation of the vacuum current.**

In the previous section we showed that QFT is not gauge invariant because the Schwinger term is not zero. In this section we will confirm the results of the previous section by calculating the change in vacuum current due to a change in the electromagnetic potential and showing that the result is not gauge invariant. We will work in 1-1D space-time. This will simplify the problem and avoid mathematical difficulties that appear when the calculations are done in normal 3-1D space-time. The main problem that we will avoid is the problem of divergent integrals. This greatly simplifies the discussion and makes the calculations straightforward. There is no problem trying to interpret and regularize divergent integrals. For the problem at hand we assume that the space dimension is in the z-direction.

In 1-1D space-time there is no magnetic field and the electric field is given in terms of the electromagnetic potential $A_z$ and $A_0$ by,

$$E = -\left(\frac{\partial A_z}{\partial t} + \frac{\partial A_0}{\partial z}\right) \tag{4.1}$$

The field operators are given by,

$$\hat{\psi}(z) = \sum_{r=-\infty}^{+\infty}\left(\hat{b}_r\varphi_{1,r}(z) + \hat{d}_r^{\dagger}\varphi_{-1,r}(z)\right); \quad \hat{\psi}^{\dagger}(z) = \sum_{r=-\infty}^{+\infty}\left(\hat{b}_r^{\dagger}\varphi_{1,r}^{\dagger}(z) + \hat{d}_r\varphi_{-1,r}^{\dagger}(z)\right) \tag{4.2}$$

where $\hat{b}_r$ ($\hat{b}_r^\dagger$) are the electron destruction(creation) operators and the $\hat{d}_r$ ($\hat{d}_r^\dagger$) are the positron destruction (creation) operators. These operators satisfy the usual anti-commutator relations,

$$\hat{b}_r\hat{b}_s^\dagger + \hat{b}_s^\dagger\hat{b}_r = \delta_{rs}; \quad \hat{d}_r\hat{d}_s^\dagger + \hat{d}_s^\dagger\hat{d}_r = \delta_{rs} \tag{4.3}$$

with all other anti-commutators equal to zero.

The quantities $\varphi_{\lambda,r}(z)$ satisfy,

$$H_0\varphi_{\lambda,r}(z) = \varepsilon_{\lambda,r}\varphi_{\lambda,r}(z) \tag{4.4}$$

where,

$$\varphi_{\lambda,r}(z) = u_{\lambda,r}e^{-ip_r z} \tag{4.5}$$

and where,

$$H_0 = \left(-i\sigma_x \frac{\partial}{\partial z} + m\sigma_z\right) \tag{4.6}$$

In the above expressions $\sigma_x$ and $\sigma_z$ are the usual Pauli matrices, 'r' is an integer, $\lambda = \pm 1$ is the sign of the energy, $p_r = 2\pi r/L$, and,

$$\varepsilon_{\lambda,r} = \lambda E_{p_r}; \; E_{p_r} = \sqrt{p_r^2 + m^2}; \; u_{\lambda,r} = N_{\lambda,r}\begin{pmatrix} 1 \\ p_r \big/ \left(\lambda E_{p_r} + m\right) \end{pmatrix}; \; N_{\lambda,r} = \sqrt{\frac{\lambda E_{p_r} + m}{2L\lambda E_{p_r}}} \tag{4.7}$$

and L is the 1-dimensional integration volume.

The $\varphi_{\lambda,r}(z)$ form an orthonormal basis set and satisfy,

$$\int_{-L/2}^{+L/2} \varphi_{\lambda,r}^\dagger(z)\varphi_{\lambda',s}(z)dz = \delta_{\lambda\lambda'}\delta_{rs} \tag{4.8}$$

The Hamiltonian operator is given by,

$$\hat{H} = \hat{H}_0 + \hat{V} \tag{4.9}$$

where

$$\hat{H}_0 = \sum_r E_{p_r}\left(b_r^\dagger b_r + d_r^\dagger d_r\right) \tag{4.10}$$

and,

$$\hat{V} = \int\left(-\hat{J}(z)A_z(z,t) + \hat{\rho}(z)A_0(z,t)\right)dz \tag{4.11}$$

where,

$$\hat{J}(z) = \frac{q}{2}\left[\psi(z)^\dagger, \sigma_x\psi(z)\right];\ \hat{\rho}(z) = \frac{q}{2}\left[\psi(z)^\dagger, \psi(z)\right] \tag{4.12}$$

Now we will work the following problem. Assume at the initial time $t_0$ the system is in the vacuum state $|0\rangle$ and the electromagnetic potential is zero. Now apply an electrromagnetic potential given by,

$$A_z = \frac{\partial\chi}{\partial z};\ \ A_0 = -\frac{\partial\chi}{\partial t} \tag{4.13}$$

where $\chi(z,t)$ is an arbitrary function which satisfies the initial conditions at $t_0$,

$$\chi(z,t_0) = 0;\quad \frac{\partial\chi(z,t_0)}{\partial t_0} = 0 \tag{4.14}$$

Now, under the action of this electromagnetic potential the initial vacuum state $|0\rangle$ is perturbed and evolves into the state $|\Omega(t)\rangle$ where $t > t_0$ and $|\Omega(t_0)\rangle = |0\rangle$. The relation between $|\Omega(t)\rangle$ and $|0\rangle$ is given by,

$$|\Omega(t)\rangle = \hat{U}(t,t_0)|0\rangle \tag{4.15}$$

where $\hat{U}(t,t_0)$ is a unitary operator. From Sakurai [16] we have,

$$\hat{U}(t,t_0) = e^{-i\hat{H}_0 t}\left(1 - i\int_{t_0}^{t}\hat{V}_I(t')dt' + O\left(q^2\right)\right)e^{+i\hat{H}_0 t_0} \tag{4.16}$$

where $O\left(q^2\right)$ means terms to the order $q^2$ or higher and,

$$\hat{V}_I(t) = e^{i\hat{H}_0 t}\hat{V}(t)e^{-i\hat{H}_0 t} = \int\left(-\hat{J}_I(z,t)A_1(z) + \hat{\rho}_I(z,t)A_0(z)\right)dz \tag{4.17}$$

where $\hat{J}_I(z,t)$ and $\hat{\rho}_I(z,t)$ are the current and charge operators in the interaction representation and are given by,

$$\hat{J}_I(z,t) = e^{i\hat{H}_0 t}\hat{J}(z)e^{-i\hat{H}_0 t}\,;\quad \hat{\rho}_I(z,t) = e^{i\hat{H}_0 t}\hat{\rho}(z)e^{-i\hat{H}_0 t} \tag{4.18}$$

Now we want to calculate the current expectation value of the state $\left|\Omega(t)\right\rangle$ at some time $t_f$. It will be convenient to let $t_f = 0$. Obviously, then, $t_0 < 0$. Use (4.15) and (4.16) along with $\hat{H}_0\left|0\right\rangle = 0$ and $\left\langle 0\right|\hat{J}(z)\left|0\right\rangle = 0$ to obtain for the first order change in the current,

$$\delta J_{vac}(z) = \left\langle\Omega(0)\right|\hat{J}(z)\left|\Omega(0)\right\rangle = -i\left\langle 0\right|\left[\hat{J}(z), \int_{t_0}^{0}\hat{V}_I(t)dt\right]\left|0\right\rangle \tag{4.19}$$

Now before proceeding with this calculation what can we conclude about $\delta J_{vac}(z)$ based on physical requirements of the theory? If we substitute (4.13) into (4.1) we see that the electric field $E = 0$. Therefore the electromagnetic potential (4.13) is simply a gauge transformation from zero electric field. Therefore, if the theory is gauge invariant, then $\delta J_{vac}(z)$ should be zero. Now let us proceed with the calculation and see if this is the case.

Use (4.13) and (4.17) to obtain,

$$\int_{t_0}^{0} \hat{V}_I(t)\,dt = -\int_{t_0}^{0} dt \int \left( \hat{J}_I(z',t)\frac{\partial \chi(z',t)}{\partial z'} + \hat{\rho}_I(z',t)\frac{\partial \chi(z',t)}{\partial t} \right) dz' \tag{4.20}$$

Integrate by parts and rearrange terms to obtain,

$$\int_{t_0}^{0} \hat{V}_I(t)\,dt = \int_{t_0}^{0} dt \int \left\{ \chi(z',t)\hat{S}(z',t) - \frac{\partial}{\partial t}\left(\chi(z',t)\hat{\rho}_I(z',t)\right) \right\} dz' \tag{4.21}$$

where,

$$\hat{S}(z,t) \equiv \frac{\partial \hat{J}_I(z,t)}{\partial z} + \frac{\partial \hat{\rho}_I(z,t)}{\partial t} \tag{4.22}$$

Note that $\hat{S}(z,t) = 0$ is the continuity equation in 1-1D space-time in the interaction representation. It can easily be shown that this condition is satisfied. Therefore, use this along with the initial condition (4.14) in (4.21) to obtain,

$$\int_{t_0}^{0} \hat{V}_I(t)\,dt = -\int \chi(z',0)\hat{\rho}(z')\,dz' \tag{4.23}$$

Use this result in (4.19) to obtain,

$$\delta J_{vac}(z) = i\int \chi(z',0)\left\langle 0 \left| \left[ \hat{J}(z), \hat{\rho}(z') \right] \right| 0 \right\rangle dz' \tag{4.24}$$

In order for $\delta J_{vac}(z)$ to be zero for arbitrary $\chi(z',0)$ the quantity $\left\langle 0 \left| \left[ \hat{J}(z), \hat{\rho}(z') \right] \right| 0 \right\rangle$ must be zero. However, based on our discussion of the Schwinger term it is evident that this quantity cannot be zero. This calculation is done in the Appendix and the result is

$$\left\langle 0 \left| \hat{J}(z), \hat{\rho}(z') \right| 0 \right\rangle = -\frac{q^2}{L^2}\sum_{r,s}\left( \left( \frac{p_r}{E_{p_r}} - \frac{p_s}{E_{p_s}} \right) e^{i(p_s - p_r)(z'-z)} \right) \tag{4.25}$$

Now, in the above expressions, replace the summations by integration by taking the limit $L \to \infty$ and substituting $\sum \to \int \frac{L}{2\pi} dp$ to obtain,

$$\left\langle 0\left|\hat{J}(z),\hat{\rho}(z')\right|0\right\rangle = \frac{q^2}{\left(4\pi^2\right)}\int_{-\infty}^{+\infty}dp\int_{-\infty}^{+\infty}dp'\left(\frac{p}{E_p}-\frac{p'}{E_{p'}}\right)e^{i(p'-p)(z'-z)} \tag{4.26}$$

Use this result in (4.24) to obtain,

$$\delta J_{vac}(z) = \frac{-iq^2}{\left(4\pi^2\right)}\int\chi(z',0)\int_{-\infty}^{+\infty}dp\int_{-\infty}^{+\infty}dp'\left(\frac{p}{E_p}-\frac{p'}{E_{p'}}\right)e^{i(p'-p)(z'-z)}dz' \tag{4.27}$$

Let $\chi(k)$ be the Fourier transform of $\chi(z',0)$. Therefore we can write,

$$\delta J_{vac}(z) = \frac{-iq^2}{\left(4\pi^2\right)}\int\left(\chi(k)e^{-ikz'}dk\right)\int_{-\infty}^{+\infty}dp\int_{-\infty}^{+\infty}dp'\left(\frac{p}{E_p}-\frac{p'}{E_{p'}}\right)e^{i(p'-p)(z'-z)}dz' \tag{4.28}$$

Integrate with respect to $z'$ to obtain,

$$\delta J_{vac}(z) = \frac{-iq^2}{2\pi}\int\chi(k)dk\int_{-\infty}^{+\infty}dp\int_{-\infty}^{+\infty}dp'\left(\frac{p}{E_p}-\frac{p'}{E_{p'}}\right)\delta(p'-p-k)e^{-i(p'-p)z} \tag{4.29}$$

Next integrate with respect to $p'$ to obtain,

$$\delta J_{vac}(z) = \frac{-iq}{2\pi}\int_{-\infty}^{+\infty}\chi(k)e^{-ikz}dk\int_{-\infty}^{+\infty}I(p,k)dp \tag{4.30}$$

where,

$$I(p,k) = \left(\frac{p}{E_p}-\frac{p+k}{E_{p+k}}\right) \tag{4.31}$$

Now let us examine the integrand $I(p,k)$ as $p\to\pm\infty$. Use,

$$E_p = |p|\sqrt{1+\frac{m^2}{p^2}} \underset{p\to\pm\infty}{=} |p|\left(1+\frac{m^2}{2p^2}+O\left(\frac{1}{p^4}\right)\right) \tag{4.32}$$

Also,

$$E_{p+k} = |p+k|\sqrt{1+\frac{m^2}{(p+k)^2}} \underset{p\to\pm\infty}{=} |p+k|\left(1+\frac{m^2}{2(p+k)^2}+O\left(\frac{1}{(p+k)^4}\right)\right) \quad (4.33)$$

Use this in (4.30) to obtain.

$$I(p,k) \underset{p\to\pm\infty}{\cong} \left(\frac{m^2}{2}\right)\left(\frac{1}{|p|}-\frac{1}{|p+k|}\right) \underset{p\to\pm\infty}{\cong} \left(\frac{m^2}{2}\right)\frac{(\pm k)}{p^2} \quad (4.34)$$

Therefore the integration of $I(p,k)$ with respect to $p$ is finite and well defined. It is readily evaluated as follows,

$$\int_{-\infty}^{+\infty} I(p,k)dp \underset{R\to\infty}{=} \int_{-R}^{+R}\left(\frac{p}{E_p}-\frac{p+k}{E_{p+k}}\right)dp \underset{R\to\infty}{=} \Big|_{-R}^{+R}\left(E_p - E_{p+k}\right) \quad (4.35)$$

Use the fact that $E_p = E_{-p}$ along with (4.33) to obtain,

$$\int_{-\infty}^{+\infty} I(p,k)dp \underset{R\to\infty}{=} \left(E_R - E_{R+k}\right)-\left(E_R - E_{R-k}\right) = E_{R-k} - E_{R+k} \underset{R\to\infty}{=} -2k \quad (4.36)$$

Substitute this result in (4.30) to obtain,

$$\delta J_{vac}(z) = \frac{2iq^2}{2\pi}\int_{-\infty}^{+\infty} k\chi(k)e^{-ikz}dk = -\frac{q^2}{\pi}\frac{\partial}{\partial z}\int_{-\infty}^{+\infty}\chi(k)e^{-ikz}dk = -\frac{q^2}{\pi}\frac{\partial\chi(z,0)}{\partial z} \quad (4.37)$$

From result we see that $\delta J_{vac}(z)$ is obviously non-zero. A gauge invariant theory would require that this result is zero. Therefore the formal theory is not gauge invariant.

**5. Conclusion.**

At this point we can respond to W. Greiner's comment that "there is a common belief that some artifact of the exact mathematics is the source of the problem". We have shown that this belief is not correct. The reason non-gauge invariant terms appear in the theory is because the formal theory is not, in fact, gauge invariant. Therefore then when calculations are performed in perturbation theory it should be expected that non-gauge invariant terms should appear in the results. These terms are not due to “some artifact of the exact mathematics” but are the result of a correct mathematical calculation. This is shown in Section 4 where we calculated the change in the vacuum current due to a gauge transformation in 1-1D space-time. In 1-1D space-time we avoid the problem with divergent integrals that that would appear in normal 3-1D space-time. Therefore there is no problem with interpreting the results which where shown to not be gauge invariant.

As we have shown in order for the theory to be gauge invariant the Schwinger term must be zero. However Schwinger has shown that this term cannot be zero. The reason that this term is not zero is due to the fact that the vacuum state $|0\rangle$ is the state with the minimum free field energy. In contrast Dirac hole theory, where it has been shown that the vacuum is not a minimum energy state[13][14], has been shown to be gauge invariant [13]. This suggests that for a theory to be gauge invariant requires a proper definition of the vacuum state which allows for the existence of quantum states with less energy than the vacuum state. This aspect of the problem is discussed in greater detail in [17].

How then does quantum physics ultimately produce the correct physical result which much be gauge invariant? Since the formal theory is not gauge invariant this requires the introduction of an additional step. As has been discussed in Section 1 the results of perturbation theory must be “corrected” by removing the non-gauge invariant terms. This can be done by a number of methods as discussed in Section 1.

**Appendix**

Use (4.12) and (4.2) to obtain,

$$\hat{J}(z)|0\rangle = \frac{q}{2}\sum_{r,s}\left(\varphi^{\dagger}_{1,r}(z)\sigma_x\varphi_{-1,s}(z)\left(b^{\dagger}_r d^{\dagger}_s - d^{\dagger}_s b^{\dagger}_r\right)\right)|0\rangle + T(z) \quad (A.1)$$

where

$$T(z) \equiv \frac{q}{2}\sum_{r}\left(\varphi^{\dagger}_{-1,r}(z)\sigma_x\varphi_{-1,r}(z) - \varphi^{\dagger}_{1,r}(z)\sigma_x\varphi_{1,r}(z)\right)|0\rangle \quad (A.2)$$

Note that the summations in the above expressions are taken from $-\infty$ to $+\infty$. Use (4.5) to obtain $\varphi^{\dagger}_{\lambda,r}(z)\sigma_x\varphi_{\lambda,r}(z) = p_r / \left(L\lambda E_{p_r}\right)$. Use this in (A.2) to obtain,

$$T(z) \equiv -\frac{q}{4L}\sum_{r}\left(\frac{p_r}{E_{p_r}}\right)|0\rangle = 0 \quad (A.3)$$

This equals zero because the summation is over all integers and $p_r = -p_{-r}$ and $E_{p_r} = E_{p_{-r}}$. Similarly we can show that,

$$\langle 0|\hat{\rho}(z) = \frac{q}{2}\sum_{r,s}\langle 0|\left(\varphi^{\dagger}_{-1,r}(z)\varphi_{1,s}(z)\left(d_r b_s - b_s d_r\right)\right) + T'(z) \quad (A.4)$$

where,

$$T'(z) \equiv \langle 0|\frac{q}{2}\sum_{r}\left(\varphi^{\dagger}_{-1,r}(z)\varphi_{-1,r}(z) - \varphi^{\dagger}_{1,r}(z)\varphi_{1,r}(z)\right) \quad (A.5)$$

Use $\varphi^{\dagger}_{\lambda,r}(z)\varphi_{\lambda,r}(z) = 1/L$ to show that $T'(z) = 0$.

Use the above results to obtain,

$$\langle 0|\hat{\rho}(z')\hat{J}(z)|0\rangle = \frac{q^2}{4}\sum_{r',s',r,s}\left(\varphi^{\dagger}_{-1,r'}(z')\varphi_{1,s'}(z')\varphi^{\dagger}_{1,r}(z)\sigma_x\varphi_{-1,s}(z)\langle 0|\begin{pmatrix} d_{r'}b_{s'} \\ -b_{s'}d_{r'} \end{pmatrix}\begin{pmatrix} b^{\dagger}_r d^{\dagger}_s \\ -d^{\dagger}_s b^{\dagger}_r \end{pmatrix}|0\rangle\right)$$

(A.6)

which yields,

$$\left\langle 0\left|\hat{\rho}\left(z'\right)\hat{J}\left(z\right)\right|0\right\rangle = q^2\sum_{r,s}\left(\varphi_{-1,s}^{\dagger}\left(z'\right)\varphi_{1,r}\left(z'\right)\varphi_{1,r}^{\dagger}\left(z\right)\sigma_x\varphi_{-1,s}\left(z\right)\right) \tag{A.7}$$

Use (4.5) in the above to obtain,

$$\left\langle 0\left|\hat{\rho}\left(z'\right)\hat{J}\left(z\right)\right|0\right\rangle = \frac{q^2}{4L^2}\sum_{r,s}\left(\left(\frac{\left(E_{p_s}-m\right)\left(E_{p_r}+m\right)-p_sp_r}{E_{p_s}E_{p_r}}\right)\left(\frac{E_{p_r}-m}{p_r}-\frac{E_{p_s}+m}{p_s}\right)e^{i\left(p_s-p_r\right)\left(z'-z\right)}\right) \tag{A.8}$$

After some algebra this becomes,

$$\left\langle 0\left|\hat{\rho}\left(z'\right)\hat{J}\left(z\right)\right|0\right\rangle = \frac{q^2}{2L^2}\sum_{r,s}\left(\left(\frac{p_r}{E_{p_r}}-\frac{p_s}{E_{p_s}}\right)e^{i\left(p_s-p_r\right)\left(z'-z\right)}\right) \tag{A.9}$$

In the above expression we can exchange the dummy indices to obtain,

$$\left\langle 0\left|\hat{\rho}\left(z'\right)\hat{J}\left(z\right)\right|0\right\rangle = -\frac{q^2}{2L^2}\sum_{r,s}\left(\left(\frac{p_r}{E_{p_r}}-\frac{p_s}{E_{p_s}}\right)e^{-i\left(p_s-p_r\right)\left(z'-z\right)}\right) \tag{A.10}$$

Next use $\left\langle 0\left|\hat{J}\left(z\right)\hat{\rho}\left(z'\right)\right|0\right\rangle = \left\langle 0\left|\hat{\rho}\left(z'\right)\hat{J}\left(z\right)\right|0\right\rangle^{\dagger}$ to obtain,

$$\left\langle 0\left|\hat{J}\left(z\right)\hat{\rho}\left(z'\right)\right|0\right\rangle = -\frac{q^2}{2L^2}\sum_{r,s}\left(\left(\frac{p_r}{E_{p_r}}-\frac{p_s}{E_{p_s}}\right)e^{i\left(p_s-p_r\right)\left(z'-z\right)}\right) \tag{A.11}$$

Therefore,

$$\left\langle 0\left|\hat{J}\left(z\right),\hat{\rho}\left(z'\right)\right|0\right\rangle = -\frac{q^2}{L^2}\sum_{r,s}\left(\left(\frac{p_r}{E_{p_r}}-\frac{p_s}{E_{p_s}}\right)e^{i\left(p_s-p_r\right)\left(z'-z\right)}\right) \tag{A.12}$$